\documentclass[twocolumn,showpacs,preprintnumbers,amsmath,amssymb,10pt,aps,pra,superscriptaddress,showkeys]{revtex4-1}
\usepackage{hyperref}
\usepackage{graphicx}
\usepackage{dcolumn}
\usepackage{bm}
\usepackage[T1]{fontenc}
\begin{document}

\preprint{Submitted to: ACTA PHYSICA POLONICA A}

\title{Some exact results for the zero-bandwidth extended Hubbard model with intersite charge and magnetic interactions}
\author{Konrad Jerzy Kapcia}%
\email[corresponding author; e-mail:]{konrad.kapcia@amu.edu.pl}
\affiliation{%
Electron States of Solids Division,
Faculty of Physics, Adam Mickiewicz University in Pozna\'n, Umultowska 85, 61-614 Pozna\'n, Poland
}%
\author{Waldemar K\l{}obus}
\affiliation{%
Quantum Electronics Division,
Faculty of Physics, Adam Mickiewicz University in Pozna\'n, Umultowska 85, 61-614 Pozna\'n, Poland
}
\author{Stanis\l{}aw Robaszkiewicz}%
\affiliation{%
Electron States of Solids Division,
Faculty of Physics, Adam Mickiewicz University in Pozna\'n, Umultowska 85, 61-614 Pozna\'n, Poland
}%

\date{April 2, 2015}

\begin{abstract}
The extended Hubbard model in the zero-bandwidth limit is studied.
The effective Hamiltonian consists of (i)~on-site $U$ interaction and  intersite (ii)~density-density interaction $W$ and (iii)~Ising-like magnetic exchange interaction $J$ (between the nearest-neighbors).
We present rigorous (and analytical) results obtained within the transfer-matrix method for 1D-chain in two particular cases:
(a)~\mbox{$W=0$} and \mbox{$n=1$};
(b)~\mbox{$U\rightarrow+\infty$} and \mbox{$n=1/2$} (\mbox{$W\neq 0$}, \mbox{$J\neq 0$}).
We obtain the exact formulas for the partition functions which enables to calculate thermodynamic properties such as entropy, specific heat ($c$), and double occupancy per site.
In both cases the system exhibits an interesting temperature dependence of $c$ involving a~characteristic two-peak structure.
There are no phase transitions at finite temperatures and the only transitions occur in the ground state.
\end{abstract}

\maketitle


\section{Introduction}

The interplay between density-density  and  magnetic interactions is relevant to a broad range of important materials such as manganites, multiferroics, organics, and other strongly correlated electron systems \cite{MRR1990,GL2003,DHM2001,BK2008,CR2004,CR2006,K2012,K2013,N1}.

In this paper we present some exact results obtained within transfer matrix method for the zero-bandwidth extended Hubbard model with density-density and Ising-like magnetic interactions on the one dimensional chain (\mbox{$d=1$}). The 1D-Hamiltonian considered has a form
\begin{equation}
\label{row:hamiltonian}
\hat{H}  =\sum_{i=1}^{L}{\left[ U\hat{n}_{i\uparrow}\hat{n}_{i\downarrow} + W\hat{n}_{i}\hat{n}_{i+1}
-4J\hat{s}^z_{i}\hat{s}^z_{i+1} - \mu\hat{n}_{i}\right]},
\end{equation}
where $\hat{c}^{+}_{i\sigma}$ denotes the creation operator of an electron with spin $\sigma$ (\mbox{$\sigma=\uparrow,\downarrow$}) at site $i$,
\mbox{$\hat{n}_{i\sigma}=\hat{c}^{+}_{i\sigma}\hat{c}_{i\sigma}$},
\mbox{$\hat{n}_{i}=\sum_{\sigma}{\hat{n}_{i\sigma}}$}, and \mbox{$\hat{s}_i^z=(1/2)(\hat{n}_{i\uparrow}-\hat{n}_{i\downarrow})$}. \mbox{$i+1$} is the nearest neighbor of the $i$-site in the chosen  direction (from two possible directions in a chain).
We assume the periodic boundary conditions,
i.e. \mbox{$n_{L+1 \sigma} = n_{1 \sigma}$}, where $L$ is a number of sites in the chain and \mbox{$n_{i\sigma}=\langle \hat{n}_{i\sigma} \rangle$}.
\mbox{$J_0=zJ$},
where \mbox{$z=2$} is a number of the nearest neighbors.

All the terms of Hamiltonian (\ref{row:hamiltonian}) commute with one another and are diagonal in the representation of occupancy numbers. It is
convenient to use the transfer matrix method \cite{NM1953} to find the grand partition function~$Z$.

Hamiltonian (\ref{row:hamiltonian}) can be treated as a simple effective model of insulators, in which
interactions $U$, $W$ and $J$
are assumed to include all the possible contributions and renormalizations.
Notice that ferromagnetic (\mbox{$J>0$}) interactions are simply mapped onto the antiferromagnetic ones (\mbox{$J<0$}) by redefining the spin direction on one sublattice in lattices decomposed into two interpenetrating sublattices. Thus, we restrict ourselves to a case of \mbox{$J>0$}.

Exact solutions of model~(\ref{row:hamiltonian}) for some particular cases have been obtained for the one-dimensional case (\mbox{$T\geq0$}) employing the method based on the equations of motion and Green's function formalism~\cite{MM2008,MPS2012,MPS2013} or the transfer-matrix method~\cite{MPS2011,B1971,BP1974,TK1974}. Extensive mean-field studies (exact result in \mbox{$d\rightarrow+\infty$})  \cite{KKR2010,MRC1984,R1979,R1975,R1973,KKR2012,KR2011,KR2011b,KKR2010a,KR2012}
and some Monte Carlo simulations (\mbox{$d=2$}) \cite{MKPR2012,MKPR2014,N2}
of model~(\ref{row:hamiltonian}) have been also performed.
Moreover, the exact ground state ($T=0$) results have been found for $ 2\leq d <+\infty$ \cite{N3,N4,N5,N6}.

We present rigorous results for partition functions obtained within the transfer-matrix method for one-dimensional model (\ref{row:hamiltonian}) in two particular cases:
(a)~\mbox{$W=0$} and \mbox{$n=1$};  
(b)~\mbox{$U\rightarrow+\infty$}  and \mbox{$n=1/2$} (\mbox{$W\neq 0$}, \mbox{$J\neq 0$}). 

\begin{figure*}		
\includegraphics[width=0.93\textwidth]{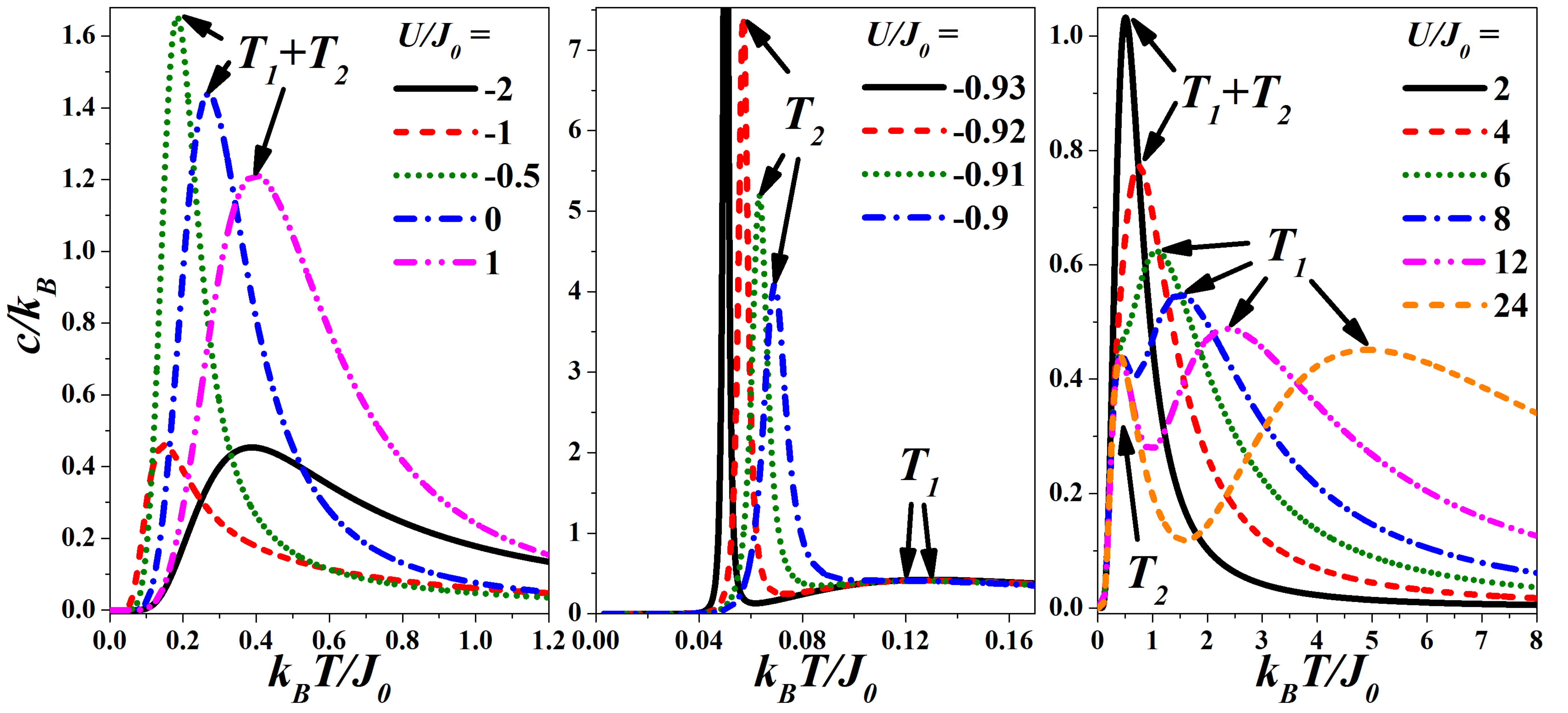}
\caption{The specific heat $c$ as a function of ${k_BT}/{J_0}$ for several values of ${U}/{J_0}$ (as labeled); \mbox{$n=1$}, \mbox{$W=0$}.}
	\label{fig:C1}
\end{figure*}

\section{Results and discussion}




(a)
For the case of \mbox{$W=0$}, a typical element of the transfer matrix for model (\ref{row:hamiltonian}) is defined as
\begin{eqnarray}
&  & P_{i,i+1} \equiv \langle n_{i\uparrow}n_{i\downarrow} | P | n_{i+1\uparrow}n_{i+1\downarrow} \rangle = \\
&=& \exp \left\{-\beta\left[(U/2) \left( n_{i\uparrow}n_{i\downarrow} + n_{i+1\uparrow}n_{i+1\downarrow}\right)  - (\mu/2) n_i  \right.\right. \nonumber\\
& - & \left.\left. (\mu/2) n_{i+1} -J(n_{i\uparrow} - n_{i\downarrow})(n_{i+1\uparrow} - n_{i+1\downarrow})\right]\right\}, \nonumber
\end{eqnarray}
where $|n_{i\uparrow}n_{i\downarrow}\rangle \in \{ |00\rangle$, $|01\rangle$, $|10\rangle$, $|11\rangle\}$ denotes a single-site state at site $i$, $\beta = 1/(k_BT)$ is the inverse temperature and $k_B$ is the Boltzmann constant.
One obtains $16$ matrix elements and the problem is reduced to diagonalization of the matrix $\check{P}$ of the form
\begin{equation}
\check{P}=\left(
\begin{array}{cccc}
  1 & x_0 & x_0 & u_0x^2_0 \\
  x_0 & m_0x^2_0 & m^{-1}_0x^2_0 & u_0x^3_0 \\
  x_0 & m^{-1}_0x^2_0 & m_0x^2_0 & u_0x^3_0 \\
  u_0x^2_0 & u_0x^3_0 & u_0x^3_0 & u^2_0x^4_0 \\
\end{array}
\right),
\end{equation}
where \mbox{$x_0 = \exp (\beta\mu/2)$}, \mbox{$u_0 = \exp (-\beta U/2)$} and \mbox{$m_0 = \exp (\beta J_0/2)$}. One can show that three eigenvalues of $\check{P}$  (\mbox{$\lambda_l$}, \mbox{$l = 1, 2, 3$}) are roots of a cubic equation:
\begin{eqnarray}\label{row:cubic}
& \lambda^3 &   -  \lambda^2 (1+2m_0x+ux^2) - \lambda \left\{xm^{-2}_0(1-m_0) \times \right. \nonumber\\
&  \times & \left. [x+m_0x+m^3_0x+m^2_0(2+x+2ux^2)] \right\} + \\
& + & x^2m^{-2}_0(1-m_0)^3(1+m_0)(1+ux^2) = 0, \nonumber
\end{eqnarray}
where $x = x^2_0$ and $u = u^2_0$, while \mbox{$\lambda_4 = 0$}.

So far the number $N$ of particles in the chain has not been specified. It can be done in a standard way by solving the following equation:
\mbox{$N = -\left(\partial \Omega/\partial \mu\right)_{\beta}$},
where $\Omega$ is the grand canonical potential, \mbox{$\Omega = -k_BT\ln Z$}. In the thermodynamic limit \mbox{$L\rightarrow\infty$} the grand sum of states $Z$ is derived as
\mbox{$Z = \lambda^L_M$},
where $\lambda_M$ is the maximum eigenvalue of $\check{P}$ (assumed to be nondegenerate). Therefore, the equation for $N$ can be rewritten as
\mbox{$\partial \lambda_M/\partial x = n \lambda_M/ x$},
where \mbox{$n=N/L$} is electron density in a system.

In the case of half-filling (\mbox{$n=1$}), the condition for $N$ can be solved analytically for arbitrary $U$ and in such a case the chemical potential is derived as
\mbox{$\mu = U/2$}
and $\lambda_M$ takes the form
\begin{equation}\label{row:Z1}
\lambda_M  =  1 + \exp{\left(\frac{\beta U}{2}\right)} \cosh{\left(\frac{\beta J_0}{2}\right)}
  +   \frac{X}{2}\exp{\left(\frac{\beta Y}{2}\right)},
\end{equation}
where
\mbox{$X = \sqrt{1 + Z_1 -4Z_2 -4Z_3 + 16 Z_4 + 4Z_5 + 2Z_6}$},
\mbox{$Y=U - J_0$},
\mbox{$Z_1 = \exp{\left(2\beta J_0\right)}$},
\mbox{$Z_2 = \exp{\left( - \beta Y/2\right)}$},
\mbox{$Z_3 = \exp{\left(\beta A /2\right)}$},
\mbox{$Z_4 = \exp{\left(\beta B /2\right)}$},
\mbox{$Z_5 = \exp{\left(- \beta Y\right)}$},
\mbox{$Z_6 = \exp{\left(\beta J_0\right)}$}, and \mbox{$A = 3J_0-U$}, \mbox{$B = 2J_0-U$}.


(b)
The limit \mbox{$U\rightarrow+\infty$} corresponds to the subspace
where the double occupancy of sites is excluded
(by electrons for \mbox{$n < 1$} or holes for \mbox{$n > 1$}).
For this case the transfer matrix elements for model (\ref{row:hamiltonian}) are defined as
\begin{eqnarray}
& & R_{i,i+1} \equiv \langle n_{i\uparrow}n_{i\downarrow} | R | n_{i+1\uparrow}n_{i+1\downarrow} \rangle = \\
& = & \exp \left\{-\beta\left[ W(n_{i\uparrow} + n_{i\downarrow})(n_{i+1\uparrow} + n_{i+1\downarrow})  \right.\right. \nonumber \\
& - & \left.\left. (\mu/2) \left(n_i +  n_{i+1}\right) - J(n_{i\uparrow} - n_{i\downarrow})(n_{i+1\uparrow} - n_{i+1\downarrow})\right]\right\}, \nonumber
\end{eqnarray}
where $|n_{i\uparrow}n_{i\downarrow}\rangle \in \{ |00\rangle$, $|01\rangle$, $|10\rangle \}$ denotes a~single-site state at site $i$ in the limit \mbox{$U \rightarrow +\infty$}.
Therefore, in this case the matrix $\check{R}$ has the following form
\begin{equation}
\check{R}=\left(
\begin{array}{ccc}
  1 & x_0 & x_0 \\
  x_0 & m_0x^2_0w_0 & m^{-1}_0x^2_0w_0 \\
  x_0 & m^{-1}_0x^2_0w_0 & m_0x^2_0w_0 \\
\end{array}
\right),
\end{equation}
where \mbox{$x_0 = \exp (\beta\mu/2)$}, \mbox{$w_0 = \exp (-\beta W_0/2)$} and \mbox{$m_0 = \exp (\beta J_0/2)$}. The eigenvalues of $\check{P}$ are roots $\lambda_l$ (\mbox{$l = 1, 2, 3$}) of the following cubic equation:
\begin{eqnarray}
\lambda^3 & - & \lambda^2 (1+2m_0xw_0) + \nonumber \\
\label{row-cubic2}
& - & \lambda x[2 - 2m_0w_0 + x w(m^{-2}_0 - m^2_0)] + \\
& + & (1-m^{-2}_0)x^2w(1+m^2_0-2m_0w^{-1}_0) = 0, \nonumber
\end{eqnarray}
where \mbox{$x = x^2_0$} and \mbox{$w = w^2_0$}.

In this case the equation for $N$
can be solved analytically for  \mbox{$n=1/2$}. One finds that \mbox{$\mu = W_0/2 - k_BT \ln \left[ 2 \cosh \left( \beta J_0/2 \right) \right]$},
and $\lambda_M$ is derived
\begin{equation}\label{row:Z2}
\lambda_M = 1 + \exp{\left(\beta W_0/4\right)} \sqrt{\textrm{sech}\left( \beta J_0/2 \right)}.
\end{equation}

\begin{figure*}
    \includegraphics[width=0.92\textwidth]{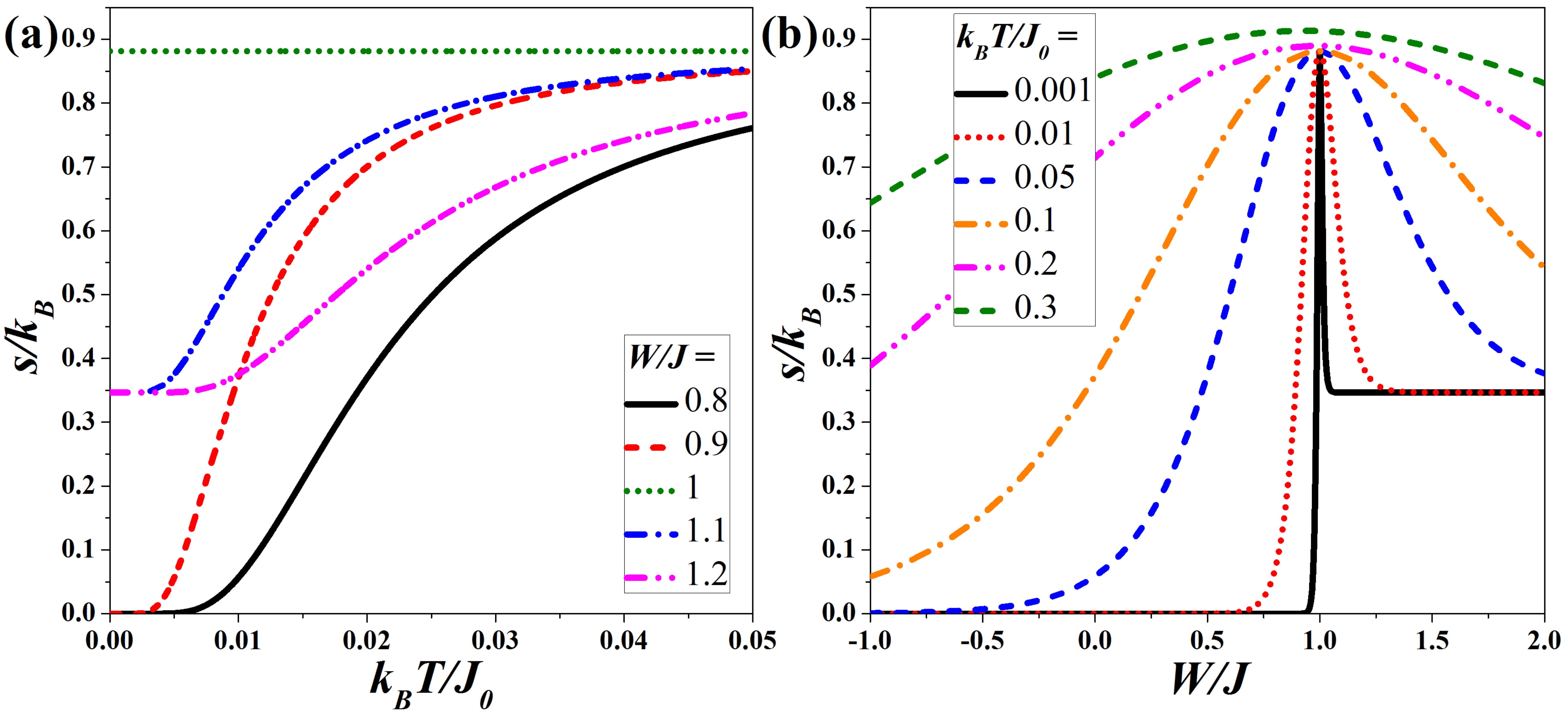}
	\caption{The entropy \mbox{$s/k_B\equiv\bar{s}$} as a function (a) of ${k_BT}/{J_0}$  and  (b) of $W/J$  (\mbox{$U\rightarrow+\infty$}, \mbox{$n=1/2$}, values of other model parameters as labeled).}
	\label{fig:SW}
\end{figure*}


The knowledge of explicit form of the sum of states $Z$ allows us to obtain thermodynamic characteristics of the system
for arbitrary temperature.
Local magnetic moment $\gamma$ is defined by:
\mbox{$\gamma  =  (1/2L)\sum_i{\langle |n_{i\uparrow}-n_{i\downarrow}| \rangle}$}.
It is related with the double occupancy $D$ per site (defined by the formula:
$D = (1/L)\left\langle \hat{n}_{i\uparrow}\hat{n}_{i\downarrow} \right\rangle = \left(\partial f/\partial U\right)_{T}$)
by the relation: \mbox{$\gamma = n/2-D$}, where
\mbox{$f = \omega + n\mu$} is the free energy of the system per site ($\omega \equiv \Omega/L$).
The entropy $s$ and the specific heat $c$ (per site) can be derived as:
\mbox{$s  =  -\partial f/\partial T$} and \mbox{$c  =  -T(\partial^2 f/\partial T^2)$}.
Because the explicit forms of the partition function $Z$ in both cases are known 
and the derivation of the above thermodynamical characteristics (i.e. $D$, $s$, $c$) is rather straightforward (\mbox{$\omega=-k_BT\ln \lambda_M$}), below we only summarize the most important conclusions following from the analysis of Eqs.~(\ref{row:Z1}) and (\ref{row:Z2}).

One can observe that the system exhibits an interesting temperature dependence of $c$ involving a~characteristic two-peak structure for some values of model parameters in the cases analyzed.
In both cases considered above there are no phase transitions at finite temperatures (in the agreement with Mermin-Wagner theorem \cite{MW1966}) and the only transitions can occur in the ground state.


\subsection{The case of \mbox{$W=0$} and \mbox{$n=1$}.} 

For large $U/J_0$ $c$~exhibit two peak structure, whereas  for \mbox{$U \lesssim 6$} peaks merge and there is a~single peak in $T$-dependence of $c$ (labeled
as \mbox{$T_1+T_2$}, cf. Fig.~\ref{fig:C1}). For \mbox{$-1<U/J_0 \lesssim -0.9$} two peaks of $c$ appear again.
The broad one (at higher temperature  $T_1$)  is connected with continuous changes in short-range charge on-site ordering (associated with $U$ term). The narrow one (at lower temperature \mbox{$T_2<T_1$})  is connected with short-range intersite magnetic ordering ($J$ term).
With decreasing of $U/J_0$ their locations move towards lower temperatures.
If \mbox{$U/J_0 < -1$} the single maximum of $c$ (connected with short-range on-site ordering) exists only and it moves toward higher temperatures with increasing of $|U|/J_0$.
In the limit \mbox{$U\rightarrow + \infty$}  (\mbox{$n=1$}) the specific heat exhibits a~single peak described by
$c^{Is} = k_B \left[(\beta J_0/2) \times \textrm{sech}\left(\beta J_0/2\right)\right]^2$,
which corresponds to 1D-Ising model in the absence of magnetic field (the peak connected with on-site ordering is ,,located'' at \mbox{$T_1\rightarrow+\infty$}).
The divergence of $c$ at \mbox{$U/J_0\rightarrow -1$} and \mbox{$T_2 \rightarrow 0$} indicates that the first-order transition occurs between the nonordered state of double occupied sites (\mbox{$\gamma=0$}) and the ferromagnetic homogeneous phase (stable for \mbox{$U/J_0 > -1$}), where all sites are singly occupied (\mbox{$\gamma=1/2$}).
One can derive the same conclusion of \mbox{$T=0$} properties of the system from a~behavior of the entropy $s$.
For \mbox{$U/J_0 > -1$} the system is magnetically ordered with \mbox{$\bar{s}(0) = 0$} (a~number of states is \mbox{$g = 2$}, \mbox{$s=k_B \bar{s} = (k_B/L)\ln g$}).
For \mbox{$U/J_0 < -1$}  the system consists of nonordered on-site electron pairs and \mbox{$\bar{s}(0) = \ln2$}  (\mbox{$g = 2^L$}).
At \mbox{$U/J_0 \rightarrow -1$} \mbox{$\bar{s}(0) = \ln2$}.
Notice that in the limit
\mbox{$T \rightarrow \infty$} the entropy \mbox{$\bar{s} \rightarrow 2\ln2$} for any $U/J_0$.

Our results for  \mbox{$W=0$} and \mbox{$n=1$} are in an agreement with the results of Ref.~\cite{MM2008} obtained using the Green's function formalism, whereas the numerical analyses of (\ref{row:cubic}) and the condition for $N$ (for arbitrary $n$ or $\mu$) should be consistent with the results of Ref.~\cite{MPS2013}.

\subsection{The case of \mbox{$n=1/2$} for \mbox{$U \rightarrow +\infty$}.} 

At \mbox{$T=0$} for \mbox{$W/J = 1$} the transition between the homogeneous charge-ordered (CO) phase (for \mbox{$W/J > 1$}) and phase separated (PS) state occurs, cf. also Ref.~\cite{KKR2012}.
For \mbox{$W/J<1$} the system is divided into two equal-sized domains: one ferromagnetically ordered completely filled by electrons  (\mbox{$n=1$}) and the other empty (\mbox{$n=0$}).
The behavior of $c$ at \mbox{$T>0$} is very similar to that discussed in the previous case (simplifying, for qualitative discussion only \mbox{$U/J_0\leftrightarrow -W/J$} replacement is needed, short-range charge order peak in $c$ is associated with $W$ term).
In the limit \mbox{$W/J \rightarrow -\infty$} $c$ exhibits a~single maximum described by the characteristic dependence for 1D-Ising model, but
in such a~case the specific heat $c^*$ of the  system is $2$ times smaller than the result $c^{Is}$  obtained in a case of \mbox{$n=1$}, \mbox{$W=0$} (\mbox{$c^*=c^{Is}/2$}).
It can be derived that at \mbox{$T=0$}: (i) for \mbox{$W/J > 1$}: \mbox{$\bar{s}(0) = (1/2)\ln2$} (the CO phase, \mbox{$g=2^{(L/2+1)}$})  and (ii) for \mbox{$W/J < 1$}: $\bar{s}(0) = 0$ (the PS:F/NO state, \mbox{$g=2L$}). If \mbox{$W = J$} the ground state is highly degenerated and \mbox{$\bar{s}(0) = \ln (1+ \sqrt{2})$} (cf. Fig.~\ref{fig:SW}). In the limit \mbox{$T \rightarrow \infty$} the entropy \mbox{$\bar{s} \rightarrow (3/2)\ln2$}.

The detailed discussion of thermodynamic properties of one-dimensional model (\ref{row:hamiltonian}) in a~general case will be the subject of a~subsequent paper.

\begin{acknowledgments}
The work (K.J.K, S.R.) has been financed by National Science Centre (NCN, Poland) as
a~research project under 
grant No. DEC-2011/01/N/ST3/00413 and a~doctoral scholarship No.  DEC-2013/08/T/ST3/00012.
K.J.K. thanks for the financial support from the ESF -- OP ``Human Capital'' -- POKL.04.01.01-00-133/09-00 -- ``\textit{Proinnowacyjne kszta\l{}cenie, kompetentna kadra, absolwenci przysz\l{}o\'sci}''.
K.J.K. and W.K. thank the Foundation of Adam Mickiewicz University in Pozna\'n for the support from its scholarship programme.
\end{acknowledgments}


\end{document}